\documentclass[11pt,twoside]{article} 
\usepackage{epsfig} 
\usepackage{amsfonts,amssymb} 
\usepackage{ihep} 

\begin{document}

\begin{center} 
\Large{\bf DO WORMHOLES FIX the COUPLING CONSTANTS?} \\ 
\vspace{0.25in} 
 \normalsize{\bf S.G. Goradia} 
\vspace{0.10in} 
 
{\it Gravity Research Institute, Inc. 
 South Bend, Indiana 46637, USA\\ 
sg@gravityresearchinstitute.org} 
\end{center} 
\date{} 
\medskip

 
\begin{abstract} 
If Newtonian gravitation is modified to use surface-to-surface 
separation between particles, it can have the strength of nuclear 
force between nucleons. This may be justified by possible existence of 
quantum wormholes in particles. All gravitational interactions would 
be between coupled wormholes, emitting graviton flux in proportional 
to particle size, allowing for the point-like treatment above. When 
the wormholes are $1$ Planck length apart, the resultant force is 
$10^{40}$ times the normal gravitational strength for 
nucleons.  \end{abstract} 
 
\medskip 
\section{Introduction} 
Newtonian gravity encounters issues for microscopic dimensions and 
cannot explain the nuclear binding force. 
%
%
%
%
Experimentalists and string theorists face a yet incomplete task of 
detecting and incorporating the spin 2 graviton into a fully quantized 
and renormalized theory. 
%
%
%
%
If we use the surface-to-surface separation between these 
particles to quantify the gravitational attraction instead of the 
center-to-center separation, 
%
%
%
at small separations 
relative to the particle radii the force between these particles grows 
much larger than classical gravity, and may resolve the above issues. 
 
\section{Modification of the Inverse Square Law} 
As an example, for two coupled nucleons (Fig. 1a), I chose the Planck 
length $L = (Gh/c^3)^{0.5}$ as the surface separation, as it is the 
minimum possible spatial distance that makes any sense in 
physics. Assuming zero separation distance would imply that the two 
particles are joined to form one particle, losing their distinctions 
as separate particles. The diameter of the nucleon is about 1 fm 
($10^{-15}$ meters). The Newtonian gravitational force is then $F_N = 
Gm^2/D^2$, where $D$ is the center-to-center distance, $\sim 1$ fm. 
If we select the surface-to-surface separation instead, the force 
would become $F_P = Gm^2/d^2$, with $d = L = 10^{-20}$ fm. 
The ratio of these two forces is $D^2/d^2 = 10^{40}$, which is also 
the strength of the proposed gravity 
relative to Newtonian gravity. 
%
%
As the nucleons are separated, $D/d$ shrinks, and $F_P$ rapidly 
approaches $F_N$~\cite{shanti}. A similar analysis can be made of the 
quark-lepton interaction (Fig. 1b). 
%
%

Nucleons are responsible for over 99 percent of gravity, therefore 
they are the primary focus of this paper. 
For nucleons, I recover Newtonian gravity at $1000$ fm. 
%
%
This modification yields a force with high intensity at short range, 
rapidly falling off to a very low intensity at long range. 

The values 
of a field and its rate of change with time are like the position and 
velocity of a particle. This modification meets the uncertainty 
principle requirement that the field can never be measured to be 
precisely zero. 

Einstein, in a paper written in 1919, attempted to demonstrate that 
his gravitational fields play an important role in the structure and 
stability of elementary particles. His hypothesis was not accepted 
because of gravity's extreme weakness~\cite{shrivatsava}. While 
Einstein's attempt is worth mentioning, it is not the foundation of 
my theory. Einstein could be wrong, but it seems he may not be. It 
has been proposed that the gravitational constant inside a hadron is 
very large, $\sim 10^{38}$ times the Newtonian 
$G$~\cite{shrivatsava}. This ``strong gravity'' inside the hadron is 
similar to my proposed modification, but in my modification, instead 
of needing to change $G$ itself, I change the distance measurement 
and get the same result. My theory does not create a conflict with 
the color force theory either. Strong gravity is consistent with 
string theory~\cite{shanti}. The short range forces are weakened at 
long range by a high order of magnitude. This makes other attributes 
of the short range forces, infinitesimal at long range.

\begin{figure}[ht] 
\centerline{\psfig{file=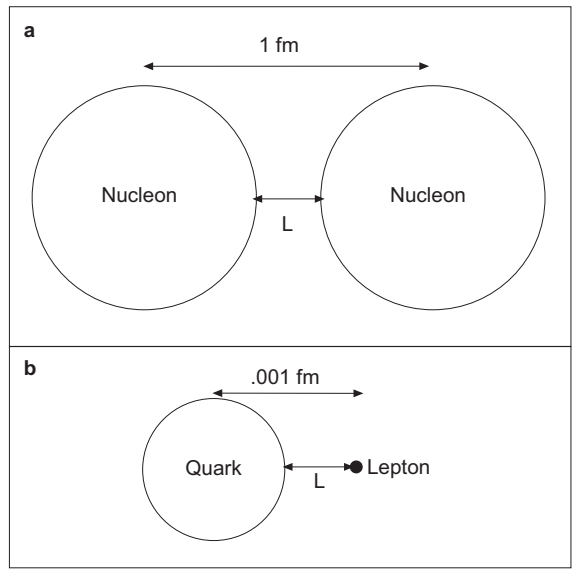,width=7.5cm}} \vspace*{8pt} 
\label{figure1} 
\caption{\small Pictorial view of gravitational interaction 
showing surface and center separations (not to scale). $L$ is the 
Planck length, $10^{-20}$ fm. {\bf a,} Two nucleons at minimum 
separation; {\bf b,} A quark and a lepton, also at minimum 
separation. The standard inverse-square law would use the 
center-to-center distances to calculate the force between the 
particles; using the surface-to-surface distance yields a much 
stronger force for these separations, equal to the relative 
strengths of the strong and weak nuclear forces, respectively.} 
\end{figure} 
 
One may question the mathematically simple application of the Planck 
scale to a problem where the relevant distances seem to be fm. Frank 
Wilczek has written a series of articles~\cite{frank}, explaining how 
these scales can be reconciled and provided responses. While this 
may seem simplistic, it seems to be mathematically valid, and 
frequently significant problems can be solved simply in the end, as 
also illustrated by Morris and Thorne~\cite{morris}. Complexity in 
physics lies in the abstraction of simplicity. Classical centers of 
shapes and therefore surfaces, though used here only for intuitive 
reasoning are invoked in nuclear coupling constants by implicit 
comparison to Newtonian gravity and in other descriptions in modern 
physics. My model is very consistent and therefore suggestive, 
however it does not reconcile the fact that nucleons overlap. Thanks 
are due to Dr. G.'t Hooft for this comment. Quantum wormholes, as 
currently theorized, may resolve this issue and give a mathematical 
foundation to my model. 
%
%
%
\section{Quantum Wormhole Connection} I postulate that each 
particle is associated with a Planck length size wormhole. The 
wormhole's exit mouth then represents the entire mass of the 
particle and propagates its $1/r$ potential to the rest of the 
universe. All gravitational interactions become interactions between 
these wormholes. Radiation by particles would consist of energy 
being absorbed by one mouth of the associated wormhole and emitted 
by the other mouth. This would justify the use of point-like 
gravity. The mouth emitting the gravitational radiations does not 
have to be at the surface, allowing the nucleons to overlap. This 
may sound like a radical approach, but it is not. The direction of 
my proposal coincides with that in the particle related article by 
Einstein and Rosen~\cite{einstein}, introducing what is now known as 
Einstein--Rosen bridges. The abundance of Planck-length size 
wormholes required could have evolved from perturbations in the 
initial big-bang density. 
%
%
 
  Stable wormholes require ``exotic", negative energy matter ``... it 
is not possible to rule out the existence of such material; and 
quantum field theory gives tantalizing hints that such material 
might, if fact, be possible''~\cite{morris}. The stability of 
wormholes is on firmer grounds now. ``...the theoretical analysis of 
Lorentzian wormholes is ``merely" an extension of $known$ 
$physics$-no new physical principle or fundamentally new physical 
theories are involved~''\cite{matt}. 
%
%
Literature search reveals no detection of any central force within nucleons, 
%
%
raising a question about the existence of 
gravitons within nucleons. Fig. 2 shows the mental picture of the 
graviton flux from nucleons with some background data. 
%
%
Richard Feynman seems to have investigated transfusion of two 
particles into gravitons~\cite{acta_physics}, but not in this 
context. The structure of the quantum space-time is 
foamy~\cite{hawking}. The potential conversion of two gluons into one 
graviton and vice versa would be debatable. Such foamy structure may 
give green light for some form of particle transfusion. 

\begin{figure}[bt] 
\centerline{\psfig{file=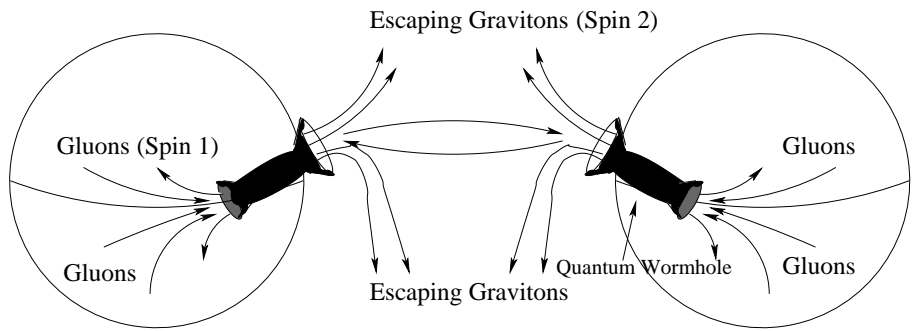,width=12cm}} 
\vspace*{8pt} 
\label{figure3} 
\caption{\small Mental image of nuclear 
interactions via quantum wormholes. The graviton flux would be 
proportional to the mass of the interacting particle, yielding 
couplings of $10^{40}$ for nucleons, $10^{34}$ for lighter 
quark-lepton pairs and $\sim 1$ for point-like leptons.} 
\end{figure} 
 
 
All long range forces are potentially simple, cumulative long range 
manifestations of their short range counter parts and vice versa 
with their intermediate range immeasurable by microscopic or 
macroscopic means. Since the spin-dependent nuclear force can be 
negative, my theory suggests investigation of photons instead of 
gravitons as the mediators of gravity. My model showing the strong 
gravity as a function of $D^2$ instead of particle mass (logical 
function of $D^3$) may point to holographic principle. Mach 
principle may imply that the universe spinning in the reference 
frame of nucleons may subject the nucleons to some form of gravity, 
not residual color force. So long as the observable characteristics 
of proposed wormholes are stable, their stability and types are of 
secondary importance because the coupling constants are averages of 
observations. The understanding of the coupling constants lies at 
the heart of our understanding other important issues. Using the 
concept of strong gravity, one can show the stability and structure 
of elementary 
particles, which could not be achieved by weak gravity~\cite{shrivatsava}. 
 
\section{Prediction} 
My model provides a consistent, intuitive and simplistic, but 
mathematical explanation of the observed relative values of coupling 
constants, something no other theory has done. Experimentally, my 
theory may be explored by a careful examination of the nuclear force 
at distances above $10$ fm. Recently published test results verified 
the gravitational inverse square law down to $218 \mu$m~\cite{hoyle}. 
The test results do not verify the higher dimensional theories that 
motivated the test, but they are not in conflict with my theory, as 
at these separations my modified force should be indistinguishable 
from Newtonian gravity. Spin-zero pions are 
%
potentially pushing the 
nucleons apart to prevent the collapse of nuclei and not pulling them 
together as theorized. Their range matching the size of nuclei gives a 
green light for such an investigation. 
 
\section{Conclusion} 
In summary, in the early part of last century, when the nuclear 
force was declared to be a separate force, the Planck length and its 
implications were not well understood. Planck's system of 
fundamental units was considered heretical until came the proposal 
by Peres and Rosen~\cite{peres}. The weakness of gravity was 
unquestioned. Therefore, it was impossible to explain strong gravity 
force in terms of Newtonian gravity and Einstein's view was 
undermined. In light of my article this issue needs to be revisited. 
My consistent results show that strong gravity creates an illusion 
of a different force between nucleons. 
%
Mathematically, the strong force coupling constant $C_s=D^2$, 
where $D=$ nucleon diameter in Planck lengths. 
\subsection*{Acknowledgments} 
I thank the staff at the Physics Department, University of Notre Dame for comments.



\begin{thebibliography}{00}
\bibitem{shanti} S. G. Goradia, physics/0210040. 
\bibitem{shrivatsava} S. K. Shrivastava, {\it Aspects of Gravitational 
Interactions} (Nova Science Publishers, Commack, 1998), p. 90. 
\bibitem{frank} F. Wilczek, {\it Physics Today} \textbf{54}, 12 (2001). 
\bibitem{morris} M. S. Morris and K. S. Thorne, {\it Am. J. Phys.} 
{\bf 56}, 395 (1988). 
\bibitem{einstein} A. Einstein and   N. Rosen, {\it Phys. Rev.} {\bf 48}, 73 (1935). 
\bibitem{matt} M. Visser, {\it Lorentzian Wormholes, From Einstein to 
Hawking} (Springer-Verlag, New York, 1996), p. 369. 
\bibitem{acta_physics} R. Feynman, \textit{Acta Phys. Pol.} {\bf 24}, 
697 (1963). 
\bibitem{hawking} S. W. Hawking, {\it Phys. Rev. \textbf{D}}{\bf 46}, 603 
(1992). 
\bibitem{hoyle} C. D. Hoyle {\it et al.}, {\it Phys. Rev. Lett.} 
{\bf 86}, 1418 (2001). 
\bibitem{peres} A. Peres and N. Rosen, {\it 
Phys. Rev.} {\bf 118}, 335 (1960). \end{thebibliography}
\end{document}